\documentclass[5p,times]{elsarticle}

\newtheorem{prop}{Proposition}

\newtheorem{example}{Example}
\newtheorem{definition}{Definition}

\usepackage{amsmath,amsfonts,amssymb}
\usepackage{ulem}
\usepackage{listings}
\begin{document}


\title{Towards the Future: Bring Program Correctness back to the focus}

\author[1]{Chongyi Yuan}{}
\author[1]{Lijie Wen}{}
\ead{wenlj@tsinghua.edu.cn}
\author[1]{Xiongliang Yan}{}

\address[1]{School of Software, Tsinghua University, Beijing {\rm 100084}, China}

\begin{abstract}
Program correctness used to be the main concern of computer software in the early days when formal semantics was a hot topic. But, the word “correct” was afterwards replaced by reliable, robust and trustworthy etc., a tradeoff situation then. This is not because correctness is no longer important, but because people found no way to get through in this direction. The tradeoff has led software engineers to focus on techniques and testing tools. Rapid development of software engineering has now reached a peak and programmers are now working freely without worrying too much about bugs, since bugs are not avoidable anyway. While software engineers are enjoying their work, computer applications are, in the name of artificial intelligence, big data and cloud computing etc., creating big fortune.

Can we always ignore existence of errors like today? Or equally, is it meaningful to talk about program correctness today? Our answer is yes. It is the time to seriously consider correctness again, before it is too late, to prepare for the future. Future generation computer systems should be correct, both syntactically (statically) and semantically (dynamically). 

The book “OESPA: Semantic Oriented Theory of Programming” (2019) by the first author has opened a new direction for semantic study. Theoretically speaking, it is possible now, based on OESPA, to compute program semantics from program text so that program correctness could be proved.

But, semantic computations and correctness proving cannot be done by hand when the size of a program is big. Automatic tools are necessary. This paper tries to lay a foundation for developing needed auto tools, so that OESPA is enriched to serve future need. To this end, a new concept named conditional semantic predicate is proposed. Concepts in OESPA, including semantic functions, semantic predicates, semantic formulas and semantic calculus, are re-represented in accordance. Such re-introduction is necessary since the book is the only publication on semantic calculus so far. The new version of semantic calculus illustrates how semantics auto-computation would be carried out. The second and third authors have played an important role in the course of this stage of study.

A 3-step meta-method called ARM is the tour-guide in writing this paper. As readers may find that the first step of ARM, i.e. abstraction of physical object (variables here), is the key step for OESPA. ARM aims at constructing formal models whenever a formal model is needed. OE is a formal computation model for programming, SPA provides a formal way in defining program semantics (A), and doing semantics predicate (SP) calculus.
\end{abstract}


\maketitle

\section{Introduction}\label{sec:intro}

Our study is of 2-in-1 nature. By ``2-in-1'' we mean two things: syntax and semantics in one model (i.e. OESPA), and physical observation and logical analysis in one meta-method (i.e. ARM). It is ARM that has made OESPA possible. ARM, abbreviated from \emph{A}bstraction, \emph{R}epresentation, \emph{M}ethod, was proposed by the first author at a CBPM (Chinese Business Process Management, 2014) meeting, but not published at that time. ARM aims at constructing a formal model for a given application. It is applicable whenever a formal system model is needed. Thus, it deserves a few words of introduction. The abstraction step in ARM requires a purpose-guided observation on physical objects involved in the given problem. Such an observation-based abstraction serves as the bases for constructing (representation step) and analyzing (method step) the wanted formal model.

Physical objects involved in programming are memory locations where data are stored and programs are executed. With semantic computing as the purpose, careful observation on memory locations has resulted as a formal definition of program variables (see Definition~\ref{def:pv} and~\ref{def:ov} below). As a formal model for programming, OESPA is a nature consequence of variable definition in which OE (\emph{O}peration \emph{E}xpression) is for programming and SPA (\emph{S}emantic \emph{P}redicate and semantic \emph{A}xioms) is for semantic definition and semantic analysis.

This paper aims at automatic computation of semantics: from understanding semantics (OESPA) to semantics computation. To this end, a new concept named conditional semantic predicate is to be proposed next with which semantic axioms are better written, and math concepts like semantic functions, semantic predicates, semantic formulas and semantic calculus are modified so that auto tools for semantic computation becomes feasible.

OESPA becomes mature from naïve in the course of the first author’s long-term teaching on formal semantics. ARM is just a by-product. It was noticed that the lack of proper abstraction of variables has prevented conventional formal semantics from proper treatment of program semantics. Next in Section~\ref{sec:history}, a brief recall is given on the way how conventional formal semantics (mainly denotational semantics~\cite{T76} , operational semantics~\cite{P81} and Hoare logic ~\cite{H69,HH98}) understand program variables and program semantics. Section~\ref{sec:hoare} introduces what Hoare logic has achieved, since Hoare logic is the main stream in this area. Other logics~\cite{D76,G81,CM88,CW96} are, almost all, followers of it.

The rest of this paper will talk about next topics:

\begin{itemize}
\item How to abstract memory locations (first step of ARM) so ambiguity would be removed from assignments, and how to form a system based on the abstractions (second step of ARM) so that semantics axioms of the system can be defined on top of its syntax formulas.
\item How to invent new methods for the analysis of the newly constructed system (last step of ARM).
\end{itemize}

In the discussion of above topics, we will talk about:

\begin{itemize}
    \item How to give a formal definition of program semantics.
    \item How to compute semantics from program text.
    \item How to develop semantics of function calls.
    \item Semantic axioms on C pointers.
\end{itemize}

\section{Program variables and program semantics in conventional formal semantics}\label{sec:history}

\subsection{Concept of program variables is copied from mathematics}

All branches of formal semantics, including denotational semantics, operational semantics and Hoare logic, have inherited the concept of variables from syntactic definition of programming languages while programming languages have copied the concept of variables from mathematics. The syntactic definition of a variable given by programming languages is just an identifier with no semantic operator(s) and operation(s) defined on it. Such a variable can be read only for evaluating math expressions that need no operator. Consequently, an assignment is not a precise math operation with explicit operator and operands, and as such, variables involved in an assignment are ambiguous: the same identifier may represent a value as well as a memory location at the same time. Such ambiguity has made formal treatment impossible.

Formal semantics aims at developing semantics from program text without evaluating expressions. As such, semantic read operations before and after an assignment (or a program segment) are a must for investigations to see what changes would be brought up by executing the assignment (or segment). In fact, such changes are, as we will see below, nothing but the semantics of the due assignment (segment). Of course, such read operations do not return a value. Instead, it returns a math expression with which the variable being read would be assigned at run time. In a word, formal semantics, denotational, operational or axiomatical, did not define the concept of program variables for semantic need.

\subsection{How to define program semantics, or what is it?}

It is often the case that different programs by different programmers are designed to carry out the same duty while they enjoy their own properties and their own ways to do it. The completion of the duty and the respective execution processes, which of them is the semantics? It depends on how you understand the word semantics.

The set of execution sequences tells a computer what to do, so this set is the semantics for a  computer. But, for a human client, the consequence of program execution is the semantics of the program, since what a client wanted is just the result. So, the semantics of $x$:=$e$ for a computer is to assign the value of $e$ (evaluated at the initial state of $x$:=$e$) to variable $x$ while for a human client, its semantics is the equation $x$:=$e$ at the final state. The problem is how to put it formally? Denotational semantics says that it is a function to map a program state to a program state, and this semantic function is its denotation. Operational semantics tells that its semantics is an operation. Denotation, and operation, they are talking about semantics for a computer, not for human clients. Hoare logic says that it is all the true assertions in the form $\{p\} x$:=$e \{q\}$. The concept of assertions will be introduced next in Section~\ref{sec:hoare}. Here, it is clear that neither denotational semantics and operational semantics, nor Hoare logic, has a direct and clear description of the informally stated semantics of $x$:=$e$. The difficulties they have encountered include: ambiguous assignments do not subject to direct semantic treatment, and a single variable name cannot denote both its initial and final values in the same formula. For example, in case the above expression $e$ is given by $x$+1, then the semantics of $x$:=$e$ for human client would become equation $x$:=$x$+1, where the first $x$is the final value while the second $x$ is the initial value. How could the same name represent two different values in the same equation?As such, conventional formal semantics are, in a sense, ``disabled'' by the miss-defined concept of variables.

Apparently, conventional formal semantics did not realize that program variables demand a definition for semantic need. OESPA starts from such a variable definition, replaces ambiguous assignment with precisely defined write/read operations. And as such, initial value and final value of the same variable can be represented, in the same formula, with the same name.

Hoare logic is the best studied branch of formal semantics. It is introduced, in certain detail in Section~\ref{sec:hoare}, to make clear the reason why it provides no essential help to software engineers.

From now on, the term “semantics” means always semantics for human client. 

\section{Hoare logic and others}\label{sec:hoare}

The most fundamental concept in Hoare logic is assertions(or a triple) on an assignment (extendable to assertions on a program) in the form $\{p\}\mathcal{S}\{q\}$ where $\mathcal{S}$ is an assignment, $p$ and $q$ are predicates on the initial state and respectively final state of $\mathcal{S}$. Assertion $\{p\}\mathcal{S}\{q\}$ is true if the truth of $p$ at the initial state of $\mathcal{S}$ leads to the truth of $q$ at the final state of $\mathcal{S}$. Compared with other semantics study branches, Hoare logic is much better, since the concept of assertions is well defined so that the semantics of an assignment may be given, by using auxiliary constants, like $\{x$=$a\}x$:=$x$+1$\{x$=$a$+1$\}$ for all constant $a$. This is much more explicit than ``an operation'', ``a denotation'' etc. Put aside the fact that assertions talk about program properties instead of program semantics, the problem is, how to prove the truth of $\{p\}\mathcal{S}\{q\}$?

The ambiguity of assignments have been wrapped up by assertions, but $\{p\}\mathcal{S}\{q\}$ is itself ambiguous: the same name in $p$ and $q$ represent its initial value and final value respectively. As such, an assertion does not comply with direct formal treatment. The development of Hoare logic was not on the track to give variables a new definition. Instead, the concept of weakest preconditions, represented as $wp(x$:=$e,q)$, was invented for the proof of $\{p\}x$:=$e\{q\}$. By definition, $wp(x$:=$e,q)$=$q^x_e$, where $q^x_e$ is obtained by substituting every appearance of $x$ in $q$ with expression $e$. Now, $p \rightarrow q^x_e$ implies $\{p\}x$:=$e\{q\}$. The most fundamental axiom in Hoare logic is nothing but $\{q^x_e\}x$:=$e\{q\}$, though written differently. Other axioms in Hoare logic are variations of this one to cover control mechanisms, including sequences, choices and loops. What has prevented Hoare logic from practical use, we believe, is the fact that it is hard to extend the above formal treatment on a single assignment to a real program. In order to prove just one property at the final state, the weakest precondition must be computed one statement by one statement starting from the last one forward. The number of statements in a real program must be great, the computing process is bound to be complicated and time-consuming. More importantly, no parallelism possible for the series computing of weakest preconditions. Besides, no one knows how many properties need to be proved in order to prove the correctness of a program. All these factors have made infeasible the automatic proving of program correctness by Hoare logic. There existed trade-offs between pure formalism and practical use: B-method and model checking are among the best. These methods had in deed attracted attentions from practitioners, but not for long.

Theoretical speaking, Hoare logic is well studied and beautiful to win Turing prize while practically speaking, it is not so sound. It doesn’t change the fact that testing is a must step in software lifecycle.

Next section is devoted to define the write/read operations. This is the first step of ARM.

To leave a deep impression on readers before starting Section~\ref{sec:pv}, let's have a comparison among operational semantics, Hoare logic and OESPA, with a small program $S$, to see how each of  which deals with the semantics of $S$.

\begin{example}A comparison by example.

Program $S$: $x:=x+y$; $y:=x-y$; $x:=x-y$.

How to study semantics of $S$, i.e., $sem(S)$?

For Operational semantics: $sem(S)=\{\langle x:=x+y, y:=x-y, x:=x-y\rangle\}$, a set consisting of one sequence only. There are three categories of properties: stable properties, safety properties and progress properties. The program $S$ given above is too small to enjoy properties in all three categories. The only meaningful property of $S$ is a progress property, namely $x=a\wedge y=b\longmapsto$ (leads-to) $x=b\wedge y=a$, which says that sooner or later $x=b\wedge y=a$ will be true if initially $x=a\wedge y=b$. This property is proved by symbolic execution (or analysis) of $S$ with $x=a\wedge y=b$ as its initial state. Termination is also a progress property of S.

For Hoare logic, to prove the same property given above, we have:

$\{wp_3\}$ $x:=x+y$; $\{wp_2\}$ $y:=x-y$; $\{wp_1\}$ $x:=x-y$ $\{q\equiv x=b\wedge y=a\}$.

By definition of weakest precondition (i.e. $wp$):

$wp_1=wp(x:=x-y, x=b\wedge y=a)\equiv x=a+b\wedge y=a$,

$wp_2=wp(y:=x-y, x=a+b\wedge y=a)\equiv x=a+b\wedge y=b$,

$wp_3=wp(x:=x+y, x=a+b\wedge y=b)\equiv x=a\wedge y=b$,

so, $p\equiv x=a\wedge y=b$, i.e., $\{x=a\wedge y=b\}S\{x=b\wedge y=a\}$.

Auxiliary constants $a$, $b$ are necessary to distinguish final value from initial value of the same variable.

OESPA:

With a precise definition of variables (names of memory locations), the concept of semantic functions is obtained. For variables $x$ and $y$, semantic functions $\uline{x}$ and $\uline{y}$ represent, respectively, the final values of $x$ and $y$ while $x$ and $y$ themselves are their respective initial values for the program in question. So, the semantics of above program $S$ is given by $\uline{x}=y\wedge\uline{y}=x$ for OESPA (see Figure~\ref{fig:sc} in Section~\ref{sec:sc} to find out how to compute it from program text).

\end{example}

Readers may find, in what follows, the following facts about OESPA:

\begin{itemize}
\item No auxiliary constants like $a$ and $b$ above, are necessary.
\item It computes, from program text, semantics rather than properties.
\item It allows parallel computing in reducing a semantic formula.
\item It contains a rich body of new math concepts including semantic functions, etc. which are necessary for semantic studies while they are missing in conventional mathematics.
\item Above characteristics ensure automatic semantics computation from program texts.
\item Key words like ``if'' and ``$\sim$'' are removed from semantic axioms with conditional semantic predicates, so that auto tools can be developed.
\end{itemize}

\section{Formal definition of program variables}\label{sec:pv}

A formal definition is a purposeful abstraction of something in doing formalism. Here we aim at semantic analysis.

\begin{definition}[Program variable]\label{def:pv}
A variable is a two-facet object. As a physical object (a memory location), it allows three operations applied on it as given by Definition~\ref{def:ov}. As a math object, it represents a value whenever it appears in a math expression.
\end{definition}

The three operations on a variable include one write operation and two read operations.  Operators chosen for them are respectively the over bar ``$\bar{\;\;}$'', the under bar ``$\_$'' and the curved under bar (tilde) ``$\uwave{\;\;}$''. Such operators would have a complete cover-up on top (or from bottom) a variable name, no matter long or short the variable name is.

Note that the write operation is the formal way to assign a new value to a variable, and as such, a new programming language can be deﬁned with assignments to be replaced by write operations and control mechanisms deﬁned accordingly. To simplify the matter, matured concepts like date types, variable declarations, function definitions and function calls etc. are not to be redefined. What to be obtained is a computation model rather than a complete programming language. This computation model is named Operation Expression, abbreviated as OE, since a program in this model is an expression constructed with the write (as well as read) operator and control mechanisms.

\begin{definition}[Operations on variables]\label{def:ov}\leavevmode\\
\indent $\bar{\;\;}$: $V\times E_m\rightarrow E_o$, write operation,\\
\indent  $\_$: $V\times E_o\rightarrow E_m$, read-after operation,\\
\indent $\uwave{\;\;}$: $V\times E_o\rightarrow E_m$, read-before operation,\\
\indent where $E_o$ and $E_m$ are respectively the set of operation expressions (to be defined) and the set of mathematical expressions on variable set $V$.
\end{definition}

These operations are represented respectively as:\\
\indent $\bar{v}(e)$: to write variable $v$ with math expression $e$,\\
\indent $\uline{v}(p)$: to read variable $v$ after (the execution of) $p$,\\
\indent \uwave{v}$(p)$: to read variable $v$ before (the execution of) $p$,\\
\indent where $p$ is an operation expression in $E_o$.

The concept of variables includes scale variables, well-deﬁned structures and elements of such structures.

\begin{example}\leavevmode\\
\indent (a) $\bar{x}(x+1)$, $\bar{A}[i](B[i])$, $\bar{A}(B)$, to write variable $x$, array element $A[i]$ and array $A$ as a whole respectively.\\
\indent (b) $\uline{x}(\bar{x}(x+y))$, $\uline{A}[j](\bar{A}[i](B[k]))$, $\uline{A}(\bar{A}(B))$, to read variable $x$, array element $A[j]$ and the whole array $A$ after respective write operations.\\
\indent (c) \uwave{x}$(\bar{v}(e))$, to read variable $x$ before operation expression $\bar{v}(e)$.
\end{example}

With a variable name fixed as the first operand and leaving the second operand absent for the two read operations, obtained are semantic functions defined on $E_o$. Semantic functions are the key to open door from program text in OE to program semantics.

\begin{definition}[Semantic function]\label{def:sf}Let $v$ be an arbitrary variable, i.e. $v\in V$. From Definition~\ref{def:ov}, we have:\\
\indent $\uline{v}: E_o\rightarrow E_m$,\\
\indent \uwave{v}: $E_o\rightarrow E_m$,\\
\indent $\uline{v}$ and \uwave{v} are called semantic functions on $E_o$. The former is abbreviated as a F-function while the latter is a I-function. 
\end{definition}

Note that when an operation expression in $E_o$ is given to the semantic functions as their respective operands, the semantic functions become complete read-operations, each to return a math expression.

Now, for every $p$,  $p\in E_o$, \uwave{v}($p$) is a math expression which would evaluate to the value of $v$ at the initial state of $p$, and $\uline{v}(p)$ is a math expression which would evaluate to the value of $v$ at the final state of $p$. The math expressions mentioned above are both defined on the initial state of $p$.

\begin{prop}[Initial and final value]\label{pro:ifv}\leavevmode\\
\indent $\uline{v}$ represents the final value of $v$ in general when no particular operation expression is in question.\\
\indent \uwave{$v$} represents the initial value of $v$ in general when no particular operation expression is in question.
\end{prop}

Proposition~\ref{pro:ifv} ensures us to represent freely how the final state relies on the initial state of a program (or program segment). Thus, there is a formal way now to describe semantics of programs. 

In contrast to conventional assignment, with the two read-operations as its inverse, the write-operation makes a big difference.

Semantic functions are used to trace back written expressions.

\begin{example}\label{exp:2}
The semantics of $\overline{x}(x+v)$ is $\uline{x}=x+v$ and $\uline{v}=v$ since by semantic axioms given blow, we have $\uline{x}(\overline{x}(x+v)) = x + v$ and $\uline{v}=v$.
\end{example}

Example~\ref{exp:2} suggests the following definition for programs in general. In Definition~\ref{pro:viv} and what follows, $S$ is always a complete program by itself. Definition~\ref{def:ps} is supported by Proposition~\ref{pro:viv} below. By “very initial value” we mean values happen to be held by variables declared, but not initialized for $S$ yet.

\begin{prop}[Very initial value]\label{pro:viv}
For program $S$, \uwave{v}(S)=v, i.e. \uwave{v}=v, for all $v\in V$.
\end{prop}

\begin{definition}[Program semantics]\label{def:ps}
The semantics of program $S$ is $\forall v\in V:\uline{v}=e_v\wedge$ \uwave{v}=v where $e_v$ is an expression which would return, when evaluated at the initial state of $S$, the final value of $v$.
\end{definition}

Remarks
\begin{itemize}
\item Assignments (write-operations), including conditional assignments and repeated assignments (loops), are the only consisting elements of any program. Thus, there must be an expression $e_v$ for variable $v$ as stated in Definition~\ref{def:ps}. 
\item Expression $e_v$ may be of special type CSP as defined next by Definition~\ref{def:csp} to serve the need of conditional assignments.
\end{itemize}

With semantic functions, we can now extend the concept of predicates to the concept of semantic predicates. A conventional predicate specifies a property of an individual object, or it describes how two or more objects are related. A program relates final values of variables with their initial values, and as such, it requires a new kind of predicates to specify variable relations (i.e. program semantics).

\begin{definition}[Semantic predicate]\label{def:sp}
The concept of semantic predicate is an extension of the concept of conventional predicate by allowing semantic functions to be its consisting part. A semantic function may appear wherever a variable appears.
\end{definition}

To be a semantic predicate, it must contain at least one semantic function. Semantic predicates are also functions defined on $E_o$. For semantic predicate $p$ and $e_o\in E_o$, $p(e_o)$ is defined as the Boole expression obtained from $p$ by substituting every semantic function $\uline{v}$ or \uwave{v} in $p$ with $\uline{v}(e_o)$ or \uwave{v}$(e_o)$.

Note that $p(e_o)$ returns a truth value, telling whether this semantic predicate $p$ is or is not a property of operation expression $e_o$.

Let $p_1\equiv \uline{x}=x+v$ and $p_2\equiv \uline{v}=v$ be two semantic predicates. For $\overline{x}(x+v) \in E_o$, we have $p_1(\uline{x}=x+v)\equiv \uline{x}(\overline{x}(x+v))=x+v\equiv x+v=x+v\equiv true$, since $\uline{x}(\overline{x}(x+v))=x+v$ (see semantics axioms in Section~\ref{sec:sfsa}). Similarly, we have $p_2(\uline{x}=x+v)\equiv \uline{v}(\overline{x}(x+v))=v\equiv v=v\equiv true$. Thus, $p_1$ and $p_2$ are properties of $\overline{x}(x+v)$.

The semantics of the write operation and the two read operations will be formally defined in Section~\ref{sec:sfsa}.

\begin{definition}[Conditional semantic predicates, CSP for short]\label{def:csp}
For semantic predicates $p_1, p_2,\cdots, p_n$ and Boole expressions $b_1, b_2,\cdots, b_n, p_1\wedge b_1\oplus p_2\wedge b_2\oplus\cdots\oplus p_n\wedge b_n$ is a conditional semantic predicate if $b_1\oplus b_2\oplus\cdots\oplus b_n$ at the initial state of the program in question. $\oplus$ is the operator exclusive-OR.
\end{definition}

As its value of $p_1\wedge b_1\oplus p_2\wedge b_2\oplus\cdots\oplus p_n\wedge b_n=p_i\wedge b_i$, we have:\\
\indent $b_i \rightarrow p_1\wedge b_1\oplus p_2\wedge b_2\oplus\cdots\oplus p_n\wedge b_n=p_i\wedge b_i = p_i$ for $i=1,2,\cdots, n$.

As an example of CSP, we have $\uline{x}=x\wedge \uline{y}=y\wedge x\leq y\oplus \uline{x}=y\wedge \uline{y}=x\wedge x>y$, which tells that $\uline{x}$ and $\uline{y}$ are ascendingly ordered. Note that $x\leq y\oplus x>y$ is always true. A conditional semantic predicate is also a function on $E_o$ that returns a conditional Boole expression, since it is not known which $b_i$ is true. 

Note that in case $n=1$, a CSP degenerates to a normal semantic predicate.

The above defined concepts, including write/read operations and semantic functions are all about individual variables. Other definitions, including CSP, are preparation for stating semantic axioms in Section~\ref{sec:sfsa}, where we approach to the second step of ARM, i.e. from individuals to a system.

{\section{Syntax formulas of OE and semantics axioms on $E_o$}\label{sec:sfsa}}

Formal syntax of OE is in fact a group of rules with which operation expressions are to be constructed. The set of so constructed operation expressions is $E_o$. OE or $E_o$, they are different as well as two sides of the same thing. Syntax and semantics, they will be given side by side. 

In formulas below, $v \in V$, $e \in E_m$ and $b$, $b_i$, $i=1, \cdots, n$ are Boole expressions.

\subsection{Assingments}

\subsubsection{Special term and simple term}\leavevmode

By simple term we mean to write a single variable, conditionally or unconditionally.

By Proposition~\ref{pro:viv} we have \uwave{v}=v for all $v \in V$, and for vaiable $u$, $u$ is not written by operation $p$, we have $\uline{u}(p)=u$ or $\uline{u}=u$. These two semantic axioms are assumed as part of all semantic axioms to be given below.

\begin{definition}\label{def:st}\leavevmode\\
\indent \textbf{Syntax formula 1}\\
\indent (F1.1) special-term $::=\varepsilon$ \\
\indent (F1.2) simple-term $::=|_v$simple-term($v$)\\
\indent simple-term($v$) $::=$unconditional-s-term($v$) $|$  conditional-s-term($v$)\\
\indent unconditional-s-term($v$) $::=\overline{v}(e)$\\
\indent conditional-s-term($v$) $::=\overline{v}(e)^b$ $|$  conditional-s-term($v$)$\odot\overline{v}(e)^b$\\
\indent \textbf{Semantic Axiom 1}\\
\indent (A1.1) $\uline{v}(\varepsilon)=v$, i.e. $sem(\varepsilon )\equiv \uline{v}=v$\\
\indent (A1.2) $\uline{v}(\overline{v}(e))=e$, i.e. $sem(\overline{v}(e))\equiv \uline{v}=e$\\
\indent (A1.3) $\uline{v}(\overline{v}(e)^b)=e$ if $b$ $\sim$ $v$ if $\neg b$, i.e. $sem(\overline{v}(e)^b)\equiv \uline{v}=e \wedge b \oplus \uline{v}=v\wedge\neg b$\\
\indent (A1.4) $\uline{v}(\overline{v}(e_1)^{b_1}\odot \overline{v}(e_2)^{b_2})=e_1$ if $b_1$ $\sim$ $e_2$ if $b_2$ $\sim$ $v$ if $\neg(b_1\vee b_2)$, \\
\indent i.e. $sem(\overline{v}(e_1)^{b_1}\odot \overline{v}(e_2)^{b_2})\equiv \uline{v}=e_1\wedge b_1 \oplus\uline{v}=e_2\wedge b_2 \oplus \uline{v}=v \wedge \neg(b_1 \vee b_2)$\\
\indent (A1.5) $sem$(conditional-s-term($v$)$\odot \overline{v}(e)^b) \equiv$
 $sem$(conditional-s-term($v$)) $\wedge \neg b \oplus sem(\overline{v}(e)) \wedge b$
\end{definition}

Remarks
\begin{itemize}
\item The symbol $|_v$ above means for every $v$, $v \in V$, exclusive-OR. 
\item A conditional-s-term($v$) allows more than one condition to be given for selecting  one value as the second operand of writing $v$. These conditions must be exclusive with each other when evaluated at runtime. Otherwise, a runtime error would be raised. Program designers are supposed to be aware of these conditions.
\item By (A1.1), the special term $\varepsilon$ writes no variable at all. It resembles ``skip''.
\item Axioms given with conditional expressions (where symbol ``$\sim$'' denotes ``or'') are included just to help understanding $sem(p)$, a new way to define axioms with CSP.
\item Note that $\odot$ can be omitted here and below. It is like the multiplication operator.
\end{itemize}

\subsubsection{Simultaneous assignment}\leavevmode

By simultaneous assignment we mean to write more than one different variable simultaneously.

\begin{definition}\label{def:simt}\leavevmode\\
\indent \textbf{Syntax formula 2}\\
\indent (F2) simultaneous-term $::=$ simple-term$\odot$ simple-term \\
$|$  simultaneous-term$\odot$simple-term $|$ (simultaneous-term)$^b$ \\
\indent \textbf{Semantic Axiom 2}\\
\indent For simple-term or simultaneous-term $p$ and $q$, let $V_p$ , $V_q$ be sets of variables written by $p$ or $q$ respectively. It is required that $V_p \cap V_q = \emptyset$.\\
\indent (A2.1) $\uline{v}(p \odot q)= \uline{v}(p)$ if $v\in V_p$ $\sim$ $\uline{v}(q)$ if $v\in V_q$ $\sim$ $v$ if $v\notin V_p\cup V_q$ \\
\indent or $sem(p \odot q)\equiv \forall v \in V: \uline{v}=\uline{v}(p)\wedge v\in V_p \oplus \uline{v}=\uline{v}(q)\wedge v\in V_q \oplus \uline{v}=v \wedge v\notin V_p\cup V_q$\\
\indent (A2.2) $\uline{v}((p \odot q)^b)= \uline{v}(p \odot q)$ if $b$ $\sim$ $v$ if $\neg b$ \\
\indent or equivalently $(p \odot q)^b\equiv p \odot q$ if $b$ $\sim$ $\varepsilon$ if $\neg b$ \\
\indent or $sem((p \odot q)^b) \equiv sem(p \odot q) \wedge b \oplus sem(\varepsilon)\wedge\neg b$\\
\indent (A2.3) $(p^{b_1})^{b_2}=p^{b_1b_2}$ or $sem(p^{b_1})^{b_2} \equiv sem(p)\wedge(b_1\wedge b_2)\oplus sem(\varepsilon)\wedge \neg(b_1\wedge b_2)$
\end{definition}

\begin{example}\label{exp:simt}\leavevmode\\
\indent $\overline{x}(1)^{y>0}\odot\overline{x}(0)^{y=0}\odot\overline{x}(-1)^{y<0}$ is a simple term to write variable $x$ so that $x$ characterizes variable $y$. This term is the same as $\overline{x}(1)^{y>0}\overline{x}(0)^{y=0}\overline{x}(-1)^{y<0}$ since $\odot$ can be omitted. $\overline{u}(e_1)\odot\overline{v}(e_2)$, $\overline{u}(e_1)\odot(\overline{v}(e_2))^b$, $(\overline{u}(e_1)\odot\overline{v}(e_2))^b$, $(\overline{u}(e_1)\odot\overline{v}(e_2))^b\odot\overline{w}(e_3)$ are different forms of simultaneous terms.
\end{example}

Remarks
\begin{itemize}
\item Each of $p$ and $q$ may be a simple-term or a simultaneous-term by itself. It is derivable from (A2.1) that $p\odot q = q\odot p$, i.e. $\odot$ is commutative.
\item It is forbidden that $p$ and $q$ write on the same variable unless the written expression evaluate, at runtime, to the same value. 
\end{itemize}

\subsubsection{Sequential control}\leavevmode

The sequential control operator is semicolon “;”, as used by many of a programming language.

\begin{definition}\label{def:sc}\leavevmode\\
\indent \textbf{Syntax formula 3}\\
\indent (F3) simple-sequence $::=$ term$;$term $|$ term$;$simple-sequence $|$  simple-sequence$;$term $|$ (simple-sequence)$^b$ \\
$|$ simple-sequence$;$simple-sequence\\
\indent term $::=$ simultaneous-term $|$ simple-term\\
\indent \textbf{Semantic Axiom 3}\\
\indent (A3.1) $\uline{v}(p;q)=\uline{v}(q)$ with \uwave{x}(q)$=\uline{x}(p)$ for all $x\in V$ or $sem(p;q)\equiv sem(q)\wedge\forall x:$\uwave{x}$(q)=\uline{x}(p) |_{del}\forall x:$\uwave{x}$(q)=\uline{x}(p)$, or $sem(p;q)=sem(q)|\forall x:$\uwave{x}$(q)=\uline{x}(p)$\\
\indent (A3.2) \uwave{v}$(p;q)=$\uwave{v}$(p)=v$ for all $v\in V$\\
\indent (A3.3) $p;q;r\equiv (p;q);r$\\
\indent (A3.4) $\uline{v}((p;q)^b)= \uline{v}(p;q)$ if $b$ $\sim$ $v$ if $\neg b$ or $sem((p;q)^b) \equiv sem(p;q) \wedge b \oplus sem(\varepsilon)\wedge\neg b$\\
\indent (A3.5) $((p;q)^{b_1})^{b_2}=(p;q)^{b_1\wedge b_2}$ or $sem((p;q)^{b_1})^{b_2} \equiv sem(p;q)\wedge(b_1\wedge b_2)\oplus sem(\varepsilon)\wedge \neg(b_1\wedge b_2)$
\end{definition}

\begin{example}\label{exp:sc}
For $p\equiv\overline{x}(x+y)$, $q\equiv\overline{y}(x-y)$, $r\equiv\overline{x}(x-y)$, to compute $sem(p; q; r)$. Since $p; q; r = (p; q); r$, let’s compute $sem(p; q)$ first. By (A1.2), $sem(q)\equiv\uline{y}=$\uwave{x}-\uwave{y}$\wedge\uline{x}=$\uwave{x}. Here \uwave{x}$\neq x$ since $q$ is not a program by itself, it is just a part of $p;q$. Similarly, we have $\uline{x}(p) = x+y\wedge\uline{y}(p)=y$. Thus, by (A3.1) we have $sem(p; q)\equiv \uline{y}= $\uwave{x}-\uwave{y}$\wedge\uline{x}=$\uwave{x}$\wedge$\uwave{x}$=x+y\wedge\uwave{y}=y\equiv\uline{y}=x\wedge\uline{x}=x+y$. Notice here the deletion of \uwave{x}$=x+y\wedge$\uwave{y}=y, denoted by $|_{del}$ in the axiom.\\
\indent Further computation tells that $sem(p; q; r)\equiv sem((p; q); r)\equiv \uline{x} = y \wedge \uline{y} = x$.\\
\indent Note that $(p;q;r)^{x>y}=(\overline{x}(x+y);\overline{y}(x-y);\overline{x}(x-y))^{x>y}$ is a conditional simple sequence. It’s easy to compute, by (A3.3), that
$sem((p; q; r)^{x>y})\equiv sem((p; q); r)\wedge x >y\oplus \uline{x} = x \wedge\uline{y} = y\wedge  x \leq y\equiv \uline{x} = y\wedge\uline{y} = x\wedge x>y \oplus \uline{x} = x \wedge\uline{y} = y \wedge x\leq y$, i.e. it sorts $x$, $y$ into ascending order.
\end{example}

Remarks
\begin{itemize}
\item ``$\uline{v}(p;q)=\uline{v}(q)$ with \uwave{x}$(q)=\uline{x}(p)$ for all $x\in V$'' denotes that the initial value of every variable $x$ in $\uline{x}(q)$ is the final value of $x$ after $p$. That is $\forall x:$\uwave{x}$(q)=\uline{x}(p)$.
\item The function of $\forall x:$\uwave{x}$(q)=\uline{x}(p)$ in (A3.1) is a value transfer: to pass the final values after $p$ to $q$ as the initial values for $q$ to start with. When this transit work is completed, $\forall x:$\uwave{x}$(q)=\uline{x}(p)$ should be deleted from $sem(p;q)$. This deleting is denoted by $|_{del}\forall x:$\uwave{x}$(q)=\uline{x}(p)$.
\end{itemize}

Proposition~\ref{pro:al} below states that the two steps of value-passing is associative since the values to be passed are final values after $p$ and after $p; q$ respectively, that are uniquely given by $p$ or by $p; q$.  

\begin{prop}\label{pro:al} For simple sequences $p$, $q$ and $r$\\
\indent $p; q; r \equiv (p; q); r \equiv p;(q; r)$  \indent\indent     \textbf{(associative law)}
\end{prop}

\subsubsection{Sequence: operation expression in $E_o$}\leavevmode

A sequence allows loop structures to be part of it. In what follows, $N$ is a constant integer no less than 0, and $\uline{b}$ is a semantic predicate, containing either F-functions only or both F-functions and I-functions. 

\begin{definition}\label{def:seq}\leavevmode\\
\indent \textbf{Syntax formula 4}\\
\indent (F4) sequence $::=$ simple-sequence $|$ loop $|$ sequence$;$sequence $|$ sequence$^b$\\
\indent loop $::=$ (sequence)$^N$ $|$ (sequence)$^{\uline{b}}$ $|$  $\varepsilon^{\uline{b}}$ $|$ term$^N$ $|$ term$^{\uline{b}}$ \\
\indent \textbf{Semantic Axiom 4}\\
\indent (A4.1) For sequence $p$, $q$, $sem(p;q)\equiv sem(p); sem(q)$\\
\indent (A4.2) $sem(p^b)\equiv sem(p)\wedge b\oplus sem(\varepsilon)\wedge \neg b$\\
\indent (A4.3) $sem(p^N) \equiv sem(\varepsilon)\wedge N$=$0\oplus sem(p)\wedge N$=$1\oplus sem(p);sem(p^{N-1})\wedge N>1$\\
\indent (A4.4) $sem1(p^{\uline{b}})\equiv sem(\varepsilon)\wedge \uline{b}(\varepsilon) \oplus(sem(p)\wedge \neg \uline{b}(\varepsilon); sem(p^{\uline{b}}))$ in case $\uline{b}$ contains only F-functions. This is the first type of repeat-until loop.\\
\indent (A4.5) $sem2(p^{\uline{b}})\equiv sem(p);sem1(p^{\uline{b}}))$ in case $\uline{b}$ contains both F-function(s) and I-function(s), it is the second type of repeat-until. \\
\indent (A4.6) $sem(\varepsilon^{\uline{b}}) \equiv sem(\varepsilon^{\uline{b}})\wedge\neg\uline{b}(\varepsilon)\oplus sem(\varepsilon)\wedge\uline{b}(\varepsilon)$ 
\end{definition}

Remarks
\begin{itemize}
\item Proposition~\ref{pro:al} above applies to sequence as well. 
\item A loop body may be just a term. That is why we have term$^N$ and term$^{\uline{b}}$ in (F4). The semantics of such a loop is covered by Semantic Axiom 4.
\item The semicolon ``;'' betweent $sem(p)$ and $sem(q)$ in (A4.1) is a semantic operator to form semantic formulas. See Definition~\ref{def:sf} in Section~\ref{sec:sc}.
\end{itemize}

\begin{example}\label{exp:seq}
For array $A[0..N-1]$, $N > 1$, \\
$p_1\equiv \overline{m}(A[0])\overline{i}(1);(\overline{m}(A[i])^{A[i]>m}\overline{i}(i+1))^{\neg(\uline{i}<N)}$ is a sequence to find the maximum element of $A: \uline{m} =max(A[0..N-1])$.\\
For array $A[0..N-1]$, $N > 1$, $p_2\equiv(q_1;q_2)^{N-1}$ is a sequence that sorts $A$ into ascending order, where $q_1$ and $q_2$ are conditional simultaneous terms: \\
\indent $q_1\equiv\bigodot_{i=0}^{N-2}(\overline{A}[i](A[i+1])\overline{A}[i+1](A[i]))^{b_1}\equiv\bigodot_{i=0}^{N-2}swap(A[i],A[i+1])^{b_1}$\\
\indent $q_2\equiv\bigodot_{i=1}^{N-2}(\overline{A}[i](A[i+1])\overline{A}[i+1](A[i]))^{b_2}\equiv\bigodot_{i=1}^{N-2}swap(A[i],A[i+1])^{b_2}$\\
\indent in which $b_1 = even(i)\wedge A[i]>A[i+1], b_2 = odd(i)\wedge A[i]>A[i+1]$ and $swap(a, b)$ is a function to exchange values of $a$ and $b$. Here the role of $swap$ is only to make clear what $q_1$ and $q_2$ do. $\bigodot_{i=1}^{N-2}$ is similar to $\prod_{i=1}^{N-2}$(product).\\
\indent To compute the semantics of a loop, the concept of loop invariant is needed. 
\end{example}

\begin{definition}\label{def:li}
For a loop with loop body $q$, the loop invariant is a semantic predicate $\uline{b}$ that is true before and after $q$. Variables that are not written by $q$ should not be counted in.
\end{definition}

Note that a loop may have more than one invariant by Definition~\ref{def:li}. But, when we talk about the loop invariant, we mean the maximum one, i.e. it implies all other invariants of this loop. 

The loop invariant of $p_1$ is $\uline{m} = max(A[0..\uline{i}-1])\wedge1\leq \uline{i}\leq N$ while $\uline{A}_M=A_M$ is not loop invariant since array $A$ is not written by the loop where $A_M$ is the multiple set consisting of all elements of $A$. So is $\uline{A}_M$. Now, $\uline{i} = N$ is the termination condition $(\neg(\uline{i}<N)\wedge 1\leq\uline{i}\leq N\equiv \uline{i} = N)$. Put the invariant and termination condition together, we have, as the semantics of $p_1$, $\uline{m}=max(A[0..\uline{N}-1])\wedge\uline{i}=N$.

The body of loop $p_2$ is $q_1;q_2$. $\uline{A}_M=A_M$ remains constant for this loop, it is part of the loop semantics as well as part of loop invariant. A loop invariant should imply progress for the loop to reach its goal. The idea behind $q_1;q_2$ is clear: to reduce the number of mis-ordered pairs in $A$. $(A[i], A[j])$, $i<j$, is a mis-ordered pair for ascending ordering if $A[i] > A[j]$. So, the loop invariant of $p_2$ is: $|\{i|\uline{A}[i]>\uline{A}[j]\wedge i<j\}| \leq |\{i|{A}[i]>{A}[j]\wedge i<j\}|\wedge \uline{A}_M=A_M$. $\uline{A}_m=A_M$ is true since $swap$ is the only operation. Loop $p_2$ will repeat its body $N-1$ times to complete the sorting since the number of mis-ordered pairs involving any individual array element is no more than $N-1$.

Loop $p_2$ provides hints to us: loop invariant is part of the semantics of loop body. For OESPA, it would be sufficient to compute semantics of loop body.  Proposition~\ref{pro:sol} given next may apply to loops in any programming language in case the loop semantics is not computed.

\begin{prop}\label{pro:sol}
If $h$ is the invariant of loop $p$ and $k$ is the termination condition of $p$, then, when $p$ terminates, we have $sem(p) = h\wedge k$.
\end{prop}

Remarks
\begin{itemize}
\item All variables appearing in semantic predicate $\uline{b}$ are constants in the loop: they represent their respective initial values of $q^{\uline{b}}$. All F-functions and I-functions in $\uline{b}$ are their respective final and initial values of each execution of loop body $q$. For conventional loops, variable names play the role of F-functions. 
\item In case $\uline{b}$ contains only F-functions, $q^{\uline{b}}$ is of repeat $q$ until $\uline{b}$ type: $q^{\uline{b}}$ ends when $\uline{b}$ becomes true. It is possible that $\uline{b}$ is true at the initial state of $q^{\uline{b}}$.
\item In case $\uline{b}$ contains both F-function(s) and I-function(s), $p^{\uline{b}}$ is often used for convergence test. For example, $\uline{b}\equiv|\uline{x}-$\uwave{x}$|\leq \delta$ is used to test how $\uline{x}$ is close to \uwave{x} after each execution of $p$ (the loop body). And as such, $p$ must be executed at least once since otherwise $\uline{x} =$\uwave{x}. $p^{\uline{b}}$ ends when$|\uline{x}-$\uwave{x}$|\leq \delta$(i.e. $\uline{b}$) becomes true. Note that \uwave{x} is not a constant: it is the initial value each time to start the loop body $p$ and it is also the final value after the previous execution of the loop body. 
\item The special loop $\varepsilon^{\uline{b}}$ is useful for constructing reactive systems. It appears before a sequence like in $\varepsilon^{\uline{b}};s$: it waits by repeated execution of $\varepsilon$ till a signal (like a button push) from outside to make $\uline{b}$ true to end $\varepsilon^{\uline{b}}$ and to start $s$.
\item We didn’t put Proposition~\ref{pro:sol} as a theorem since rigorous proof has not been found. This proposition is helpful to a programmer for him to understand his loop.
\end{itemize}

\begin{definition}\label{def:soe}
The above defined syntax formulas 1 to 4, including terms, simple sequences and sequences are the syntax of OE. Everything constructed according to this syntax is an operation expression. $E_o$ is the set of individual operation expressions.
\end{definition}

So far, what we have defined is in fact sequential OE. Definition~\ref{def:poe} tries to touch parallel OE. A parallel operation expression is semantically nondeterministic, i.e. it doesn’t have a unique final state for any given initial state. We will explain it with examples instead of defining it. An important idea is to find invariants from nondeterministic semantics. 

\begin{definition}\label{def:poe}\leavevmode\\
\indent \textbf{Syntax formula 5}, $\parallel$ is the parallel operator\\
\indent (F5) parallel-sequence $::=($sequence$\parallel$sequence$)|($parallel-sequence$\parallel$sequence$)$\\
\indent \textbf{Semantic axiom 5}, for sequences $p$ and $q$,\\
\indent (A5.1) $sem(p\parallel q)=sem(p)\otimes sem(q)$
\end{definition}

Remarks
\begin{itemize}
\item For $p$ and $q$ in parallel, there exist at least three situations: independent with each other; cooperating to provide the same service, or cooperating with due division of labor. For independent $p$ and $q$, $sem(p)$ and $sem(q)$ are also independent. We will discuss semantics of cooperating sequences with examples in a separate paper. 
\item The semantics of “$\otimes$” is nondeterministic. In case $p$ and $q$ are independent, $\otimes$ is the logic operator $\wedge$, i.e. $sem(p\parallel q)=sem(p)\wedge sem(q)$.
\end{itemize}
As the third step of ARM, analytic methods are proposed in Section~\ref{sec:sc}.

\section{Semantics Calculus (SC for short)}\label{sec:sc}

An operation expression in sequential OE consists of sequences (loops are also sequences) of terms connected by semicolon (the sequential operator). The semantics of terms are semantic predicates (SP for short) or conditional semantic predicates (CSP). Thus, the computation of program semantic starts from the computation of term semantics to yield a semantic formula as defined in Definition~\ref{def:sf} below. Semantic calculus to be introduced below aims to reduce a semantic formula to a SP or CSP, i.e. to deduce the semantics of the operation expression behind the semantic formula.

\begin{definition}\label{def:sf}
$p_1;p_2;\cdots;p_n$ is called a semantic formula if, for $1 \leq i \leq n, p_i$ is a SP or CSP.
\end{definition}

Semantic calculus is an invention to serve the need of semantics analysis of operation expressions. 

In the rules given below, $V$ is the set of variables in question, and $p$, $q$, $r$ are SP or CSP while  $V_p$, $V_q$, $V_r$ are subsets of $V$ consisting of variables whose corresponding semantic functions appear in $p$, $q$, $r$ respectively, e.g. $x\in V_p$ means $\uline{x}\in p$.

\begin{definition}[Semantic calculus, SC for short]\label{def:spc}
Semantic calculus consists of 5 reduction rules as given below.\\
\textbf{Rule 1 (Completion)}\\
With $com$ as the completion operator, and let be $p^c \equiv\wedge_{v\in V'}\uline{v}=v$, where $V'$=$V$-$V_p$, we have the completion rules below:\\
(a) $com(p)\equiv p\wedge p^c$ is the completion of $p$,\\
(b) $com(p;q;\cdots;r)\equiv com(p);com(q);\cdots;com(r)$ is the completion of $p; q;\cdots; r$.\\
\textbf{Rule 2 (Relay)} This rule is essentially the same as semantic axiom (A3.1).\\
$p;q\equiv q(v,$\uwave{v}$)\wedge p(\uline{v},$\uwave{v})$|$\uwave{v}\\
where $q(v,$\uwave{v}$)$ is obtained by replacing every $v$ in $q$ with \uwave{v}, and $p(\uline{v},$\uwave{v}) is obtained by replacing every $\uline{v}$ in $p$ with \uwave{v}. ``$|$\uwave{v}'' requires deletion of $p(\uline{v},$\uwave{v}) when every \uwave{v} in $q(v,$\uwave{v}) gains its value from $p(\uline{v},$\uwave{v}). See examples below.\\
\textbf{Rule 3 (Associative)}\\
$p;q;r\equiv(p;q);r$\\
\textbf{Rule 4 (Substitution)}\\
$p\equiv p'\wedge q\equiv q'\Rightarrow p;q\equiv p';q'$\\
\textbf{Rule 5 (Distribution)}\\
$p;(p_1\wedge b_1\oplus p_2\wedge b_2\oplus\cdots\oplus p_n\wedge b_n)\equiv (p;p_1)\wedge b_1\oplus (p;p_2)\wedge b_2\oplus\cdots\oplus (p;p_n)\wedge b_n$,\\
$(p_1\wedge b_1\oplus p_2\wedge b_2\oplus\cdots\oplus p_n\wedge b_n);p\equiv (p_1;p)\wedge b_1\oplus (p_2;p)\wedge b_2\oplus\cdots\oplus (p_n;p)\wedge b_n$.\\
And for completion:\\
$com(p_1\wedge b_1\oplus p_2\wedge b_2\oplus\cdots\oplus p_n\wedge b_n)\equiv com(p_1)\wedge b_1\oplus com(p_2)\wedge b_2\oplus\cdots\oplus com(p_n)\wedge b_n$.
\end{definition}

Remarks on above rules:\\
\indent (1)	 Rule 1 is to make the implicit part $p^c$ of $p$ explicit.\\
\indent (2) Rule 2 is, by the conjunction operation, to make the final values of variables after $p$ delivered to $q$ as its respective initial values.\\
\indent (3) Rule 3 states the fact that relay of deliveries is associative. Furthermore, it is provable that $p;q;r\equiv p;(q;r)$.\\
\indent (4) Rule 4 says that equivalents are substitutable with each other.\\
\indent (5) Rule 5 serves the need of symbolic computation, should CSP be encountered.

\begin{example}\label{exp:sc}
For operation expressions $p \equiv \overline{x}(x+y)$, $q\equiv\overline{y}(x-y)$, $r\equiv\overline{x}(x-y)$ as given in Example~\ref{exp:sc}, the semantics of $p; q; r$ was computed based on semantic axioms. Now, from respective semantics of terms $p$, $q$, $r$ obtained is $sem(p; q; r)\equiv\uline{x}= x + y; \uline{y}= x – y;\uline{x} = x – y$. This is a semantic formula. It will be reduced, by above given SC, to $\uline{x} = y\wedge\uline{y} = x$. Figure~\ref{fig:sc} above explains the reducing steps. This could be done automatically with symbolic computation tools (to be developed). Furthermore,
$sem(p; q; r)^{x>y}\equiv sem(p; q; r)\wedge x>y\oplus sem(\varepsilon)\wedge\neg(x>y)$, is reduced, by SC, to $\uline{x} = y\wedge\uline{y} = x\wedge(x>y)\oplus\uline{x} = x\wedge\uline{y} = y\wedge \neg(x>y)$.

\begin{figure}
  \centering
 \includegraphics[width=3.6in]{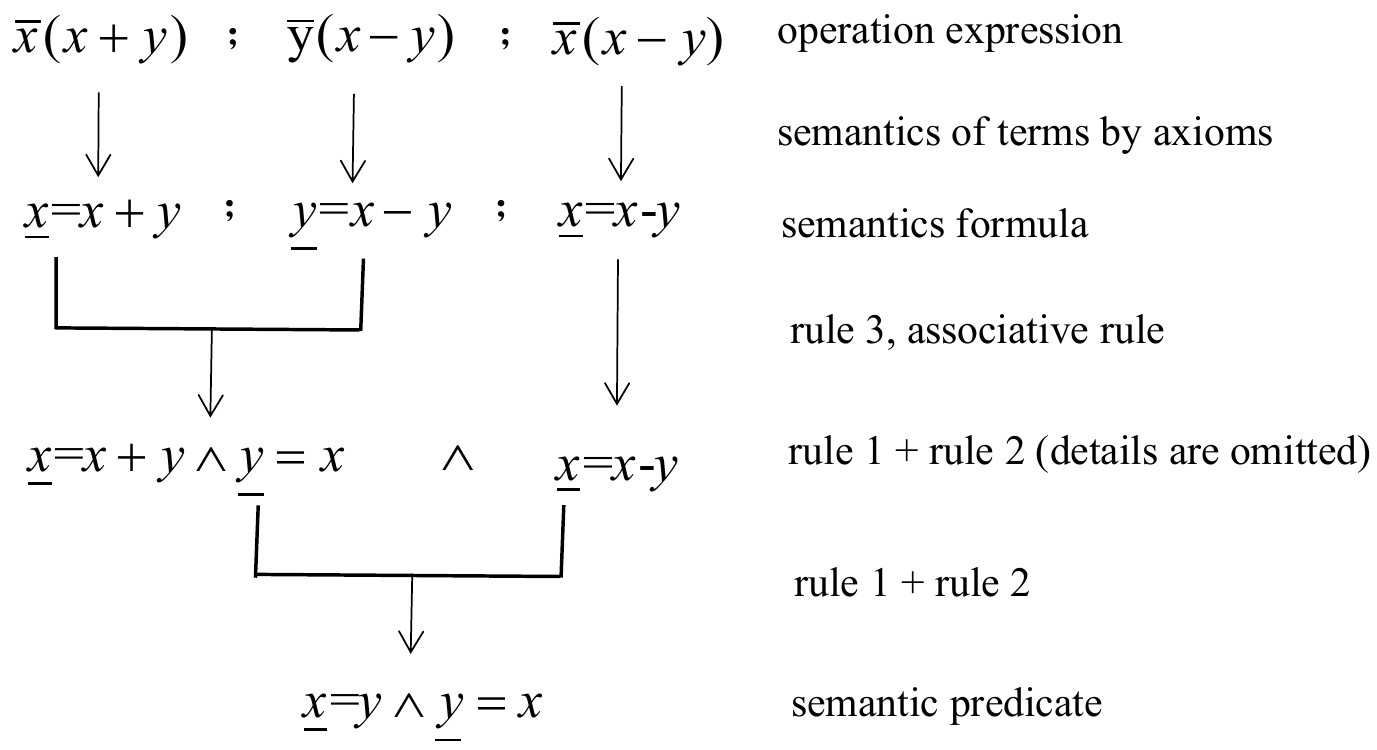}
    \caption{Semantic calculus}
  \label{fig:sc}
\end{figure}

In case $p; q; r$ is prefixed with $\overline{x}(2)\overline{y}(1)$, we have:\\
\indent\indent $sem(\overline{x}(2)\overline{y}(1)) \equiv \uline{x} = 2\wedge\uline{y} = 1$,\\
\indent\indent $sem(\overline{x}(2)\overline{y}(1); (p;q;r)^{x>y})$\\
\indent\indent $\equiv sem(\overline{x}(2)\overline{y}(1)); sem(p;q;r)^{x>y}$\\
\indent\indent $\equiv(\uline{x} = 2\wedge\uline{y} = 1);(\uline{x} = y\wedge\uline{y} = x\wedge x>y\oplus
\uline{x} = x\wedge\uline{y} = y\wedge \neg(x>y))$\\
\indent\indent $\equiv(\uline{x}=2\wedge\uline{y} = 1;\uline{x} = y\wedge\uline{y}=x\wedge x>y)\oplus
(\uline{x} = 2\wedge\uline{y}= 1;\uline{x}=x\wedge\uline{y}=y\wedge \neg(x>y))$\\
\indent\indent $\equiv\uline{x}=1\wedge\uline{y}=2\wedge 2>1\oplus
\uline{x} = 2\wedge\uline{y} = 1\wedge \neg(2>1)$\\
\indent\indent $\equiv\uline{x} = 1\wedge\uline{y}=2$.
\end{example}

This example of $\overline{x}(2)\overline{y}(1);(p; q; r)^{x>y}$ illustrates the intended symbolic computation of program semantics.

We have taken Prof. Hoare's quick-sorting~\cite{QS61} as an example in ~\cite{OESPA} to show how to develop semantics from C program text, though in a way different from this paper. We tried to find another C program of certain size as an example of computing semantics. It has turned out that every C program contains library functions. We would, to start with in the future, develop semantics of library functions. 

Now, before going on with other examples, we would like to stress once more the importance from reality to logic. Without the proper definition of program variables, OESPA would not be possible. The mate-method ARM leads a way from reality to logic. 

Next section provides two examples to illustrate the role of OESPA in understanding program features: function calls and C pointers. Table~\ref{table:com} below is a comparison between Hoare logic and our work.

\begin{table*}
\centering
\caption{A comparison between Hoare logic and OE+SP+A (OE: operation expression, SP: semantic predicate, A: axiom)}\label{table:com}
\begin{tabular}{|p{1.5cm}|p{1.5cm}|p{1.5cm}|p{1.5cm}|p{2cm}|p{1.5cm}|p{1.5cm}|p{1.5cm}|}
\hline
& \textbf{method} & \textbf{program studied} & \textbf{concept of semantics} & \textbf{concept of variables} & \textbf{core statement} & \textbf{new concept} & \textbf{process of reasoning}\\
\hline
Hoare logic & pure logic & formal syntax only & properties of programs & math objects: a value, accessible only for evaluating math expression & assignment, ambiguous & reasoning axioms based on assertions & sequential computation of weakest preconditions\\
\hline
OE+SP+A & from reality to logic & OE: formal syntax + semantic axioms & what a program does, precise definition &  a value and a physical object allowing 3 operations & write operation, not ambiguous & semantic functions predicate formula & semantic calculus, parallel computing\\
\hline
\end{tabular}
\end{table*}

\section{A study of two features from conventional programming}\label{sec:ers}

\subsection{Functions}

Function calls are often included in program texts: either called by value or called by name. Here in this subsection, we show, with examples, how to insert semantics of function calls into semantic formula. This is to mend the missing of functions in the definition of OE. Formal definitions of functions and function calls are to be discussed and proposed when OE is accepted to play a role in programming.

\begin{example}[Call by name]\label{exp:callbyname}
A program in C called quick-sorting by Prof. Hoare has been discussed in~\cite{OESPA}, of which the semantics is:

\indent $\uline{A}_M$=$A_m\wedge\forall i:0\leq i< N$-$1:\uline{A}[i]\leq\uline{A}[i$+$1]$.

The program segment\\
\indent Quick-sorting$(A);\overline{m}(A[N$-$1])$\\
\indent will assign the maximum element of $A$ into $m$. The semantics of $\overline{m}(A[N$-$1])$ is $\uline{m}$=$A[N$-$1]$, and the semantic formula for this segment is\\
\indent $\uline{A}_M$=$A_m\wedge\forall i:0\leq i<N$-$1:\uline{A}[i]\leq\uline{A}[i$+$1]$;$\uline{m}$=$A[N$-$1]$.\\
\indent That would be reduced, by semantic calculus, to\\
\indent $\uline{A}_M$=$A_m\wedge\forall i:0\leq i<N$-$1:\uline{A}[i]\leq\uline{A}[i$+$1]\wedge\uline{m}$=$\uline{A}[N$-$1]$,\\
\indent i.e., $\uline{A}_M$=$A_m\wedge\forall i:0\leq i< N:\uline{A}[i]\leq\uline{m}$.
\end{example}

This example seems naïve: it has two terms only: a call and an assignment. But on the contrary, it is meaningful in general. The way to reduce a semantic formula could be done, by semantic calculus, pairwise in parallel. For semantic formula $p_1; p_2; \cdots; p_n$, its pairwise division is $(p_1; p_2); \cdots; (p_{n-1}; p_n)$ in case $n$ is an even integer. The reduction of all these pairs is the same as shown by this example, and parallel reduction insures high efficiency.

For function calls that return a value, say $\overline{m}(\uline{sum(A)})$, where $\uline{sum(A)}$ is a function call to return the sum of all elements in array $A$. The semantics of this call is as simple as $\uline{m}$=$\sum^{n-1}_{i=0}A[i]$, since $sem(\uline{sum(A)})$=$\sum^{n-1}_{i=0}A[i]$.

\begin{example}[Recursion]\label{exp:recursion}
To find factorial of non-negative integers, i.e. for $N$, $N\geq1$, to find $N!$. The recursive program to implement function $F(N)=N!$ is given, in OE, as below, where superscript $N>1$ is a Boole expression:\\
\indent $\overline{F(1)}(1);\overline{F(N)}(N\times\uline{F(N-1)})^{N>1}$.

Note that it requires to compute $\uline{F(N-1)}$ in order to compute $\uline{F(N)}$. The semantics of above function is, by conditional semantic predicate (CSP), as below:\\
\indent $\uline{F(N)}=1\wedge N=1\oplus\uline{F(N)}=N\times\uline{F(N-1)}\wedge N>1$.\\
\indent This is abbreviated as:\\
\indent $\uline{F(N)}=(1\wedge N=1\oplus N\times\uline{F(N-1)}\wedge N>1)$.

By math induction and assuming\\
\indent $\uline{F(N-1)}=(N$-$1)\times(N$-$2)\times\cdots\times 2\times 1\wedge N>1$,\\
\indent we have\\
\indent $\uline{F(N)}=(1\wedge N=1\oplus N\times(N$-$1)
\times(N$-$2)\times\cdots\times 2\times 1\wedge N>1)$,\\
\indent i.e., $\uline{F(N)}=N\times(N$-$1)
\times(N$-$2)\times\cdots\times 2\times 1\wedge N\geq 1$,\\
\indent i.e., $N!=N\times(N$-$1)
\times(N$-$2)\times\cdots\times 2\times 1$ for $N\geq 1$.

In addition to this, 0!=1 is defined to be true. Thus, we have\\
\indent $N!=(1\wedge N=0\oplus N\times(N$-$1)
\times(N$-$2)\times\cdots\times 2\times 1\wedge N\geq 1)$.
\end{example}

This example tells that automatic tools for semantics computing must be intelligent to make use of matured math methods like math induction etc.

Another recursive function is to compute Fibonacci sequence defined by\\
\indent $Fib(0)$=0, $Fib(1)$=1 and for $n>1$, $Fib(n)$=$Fib(n$-$1)$+$Fib(n$-$2)$,\\
or in terms of CSP (in abbreviated form):\\
\indent $\uline{Fib(N)}$=$(0\wedge N$=$0\oplus 1\wedge N$=$1\oplus\uline{Fib(N-1)}+\uline{Fib(N-2)}\wedge N\geq 2)$.

The recursive OE program to produce the first $N$+2 elements of the above defined Fibonacci sequence is:\\
$\overline{Fib(0)}(0)\overline{Fib(1)}(1);\overline{Fib(N)}(\uline{Fib(N-1)}$+$\uline{Fib(N-2)})^{N>1}$.

It is a well-known fact that a recursive algorithm can always be implemented by a loop with higher efficiency. An OE program with loop to produce Fibonacci sequence is:\\
\indent $\overline{i}(0)\overline{j}(1);(\overline{i}(i+j);\overline{j}(i+j))^{\uline{i}=\uline{j}}$,\\
\indent of which the loop body is $\overline{i}(i+j);\overline{j}(i+j)$. Note that this loop will never terminate (to reflect infinite length of Fibonacci sequence) since its termination condition is $\uline{i}=\uline{j}$ that always evaluates to false. The semantics of this loop body is $\uline{i}$=$i$+$j;\uline{j}$=$i$+$j$, or by semantic calculus,\\
\indent $\uline{i}$=$i$+$j\wedge \uline{j}$=$i$+$2j$.

So, the semantic formula of the whole program is of infinite length as below:\\
\indent $\uline{i}$=$0\wedge\uline{j}$=$1;\uline{i}$=$i$+$j\wedge\uline{j}$=$i$+$2j;\uline{i}$=$i$+$j\wedge\uline{j}$=$i$+$2j\cdots$.

Thus, the above program is in dead to produce Fibonacci sequence. Different from the original recursive definition, this infinitely long formula allows parallel computing of two adjacent Fibonacci numbers, namely $Fib(2N)$ and $Fib(2N$+$1)$ for $N\geq 1$:

\indent $Fib(2N)$=$Fib(2N$-$2)$+$Fib(2N$-$1)$,\\
\indent $Fib(2N$+$1)$=$Fib(2N$-$2)$+$2Fib(2N$-$1)$.

\begin{example}[Iteration]\label{exp:iteration}
Next is a simple iterative program in C to find out root of a given equation.

\begin{lstlisting}
x0=a;  /* a is an initial approximate value */
do{
   x1=x0;
   x0=g(x1); /* compute next approximate value */
}while(fabs(x0-x1)>delta); /* delta is the allowed difference */
printf(``the root of the equation is %f'', x0);
\end{lstlisting}

To put this iteration in OE, we have:\\
\indent $\overline{x}(a);(\overline{x}(g(x)))^{\uline{b}}$,\\
\indent where $\uline{b}\equiv|\uline{x}$-$\uwave{x}|\leq\delta$ and $\uwave{x}$ is the initial value of $x$ before each execution of the loop body $\overline{x}(g(x))$. Note also that $|\uline{x}$-$\uwave{x}|$ is the absolute value of the difference. The read-before operator ``$\uwave{\;\;}$'' has made iteration much simpler.

In case the loop terminates, the semantics would be\\
\indent $\uline{x}$=$g(\uwave{x})\wedge|\uline{x}$-$\uwave{x}|\leq\delta$.\\
\indent Programmers are not responsible for termination if $g(x)$ is given to them.
\end{example}

It should be clear by now that a function call by name appears in OE programs in the same way as it appears in other programs. The quick-sorting in Example~\ref{exp:callbyname} illustrates this. A function call by value appears in OE programs is as a read-after operation applied on the function name being called like $\overline{m}(\uline{sum(A)})$. Example~\ref{exp:recursion} talked about semantics of recursive functions defined in terms of OE. Note $F(N)$ in Example~\ref{exp:recursion} is called as a whole, where $N$ is the real parameter for calling.  $F(N)$ is the operand of reading: to read the value returned by the call.

\begin{example}[Hanoi Tower]
This problem is related to an ancient Greek fairy tale. There were three poles in a row standing on the grand in a temple, with 64 disks (each has a hole in the center) of different sizes on the ﬁrst pole ($A$) while the other poles ($B$ and $C$) are empty (see Figure~\ref{fig:ht}). All disks are in order: the biggest one at the bottom and every other one is smaller than the one beneath it. It is asked to move all 64 disks from pole $A$ to pole $C$ under the conditions:

\begin{itemize}
\item[1)] One disk at a time to move from pole to pole,
\item[2)] A disk must be put on top of a bigger disk if the receiving pole is not empty,
\item[3)] Pole $B$ is the only place for a disk to stay before it ﬁnally arrives at Pole $C$.
\end{itemize}

The problem is: to make the move in the temple and to compute the move sequence here.

\begin{figure}
  \centering
 \includegraphics[width=2.8in]{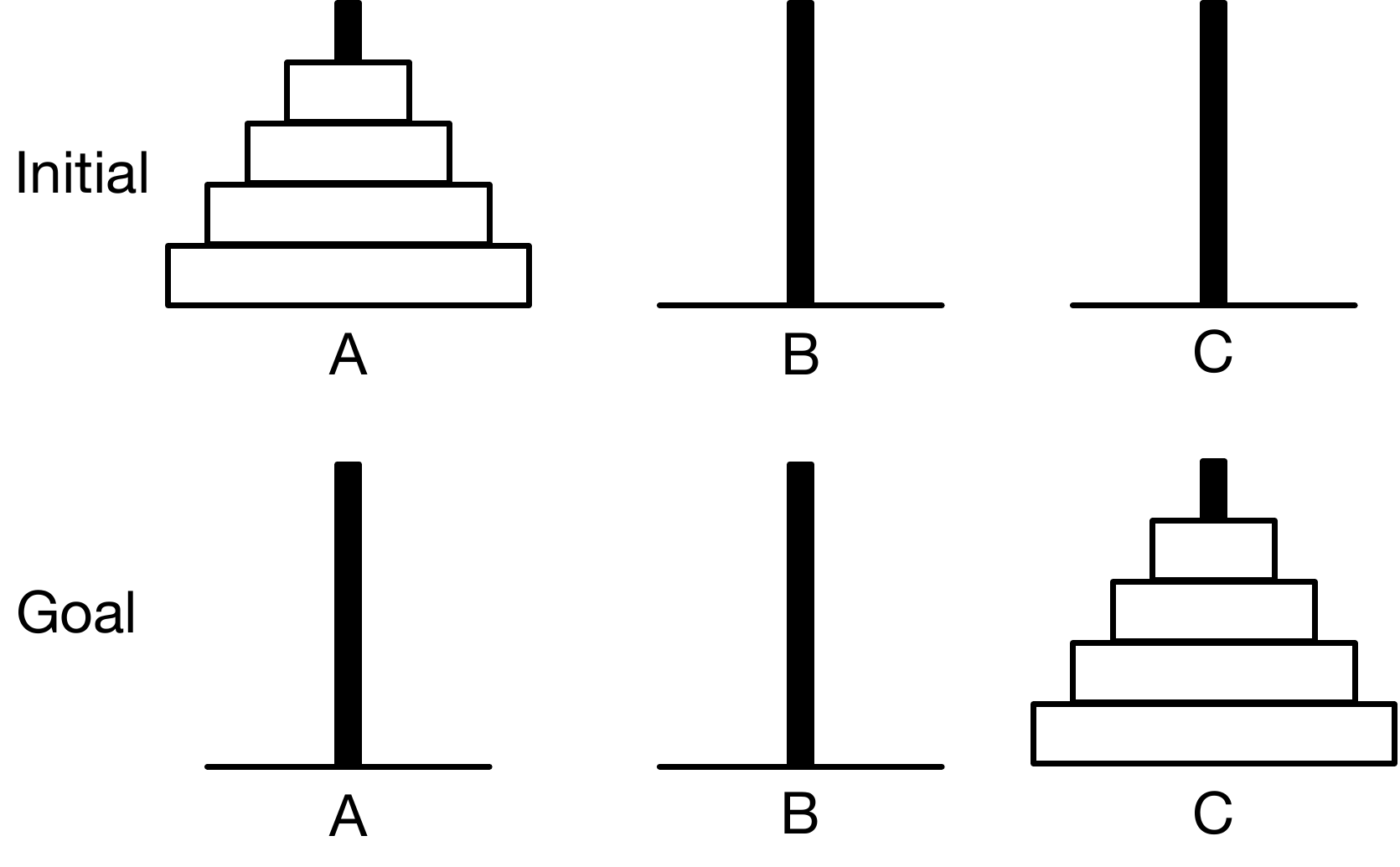}
    \caption{The initial and goal states of Hanoi Tower game}
  \label{fig:ht}
\end{figure}

For this example, it is sufficient to know that its solution is given by the following recursive function:

$T(N, A, B, C)=T(N$-$1, A, C, B); T(1, A, B, C); T(N$-$1, B, A, C)$\\
\indent and\\
\indent $T(1, A, B, C)=(A, B, C)$,\\
\indent where $N$ is an integer greater than 0. In case this function is called, it looks like $\overline{S}(\uline{T(N, A, B, C)})$ in OE: to store the sequence in $S$. This is a different way of recursion from Fibonacci and factorial since the recursive part appears twice in the definition.

The above definition of $T(N, A, B, C)$ produces the sequence for disk move: $(A, B, C)$ is to move the disk on top of pole $A$ and to put it on top of pole $C$. All moves must be done in sequence given by $T(N, A, B, C)$.

Now, semantics of $T(N, A, B, C)$, i.e. $\uline{T(N, A, B, C)}$.

We have $\uline{T(1, A, B, C)} = (A, B, C)$, and for $N > 1$,\\
\indent $\uline{T(N, A, B, C)}=\uline{T(N-1, A, C, B)}; (A, B, C);\uline{T(N-1, B, A, C)}$\\
\indent is the semantic formula. This computing goes on and terminates for given $N$. Careful observation helps to find that 
$\uline{T(N-1, A, C, B)}$ and $\uline{T(N-1, B, A, C)}$ are independent with each other, i.e. they can be computed in parallel since they are solutions to different problems. So, the above formula can be re-written as\\
\indent $\uline{T(N-1, A, C, B)}\parallel (A, B, C)\parallel \uline{T(N-1, B, A, C)}$.

This observation applies to $\uline{T(N-1, A, C, B)}$ and \\
$\uline{T(N-1, B, A, C)}$ as well. It is easy to find out that the complete sequence of disk move is a full binary tree traveled in middle order way if such analysis goes on (see Figure~\ref{fig:bt} for $N$=4). As such, the length of the sequence is $2^N$-1.

\begin{figure}
  \centering
 \includegraphics[width=3.2in]{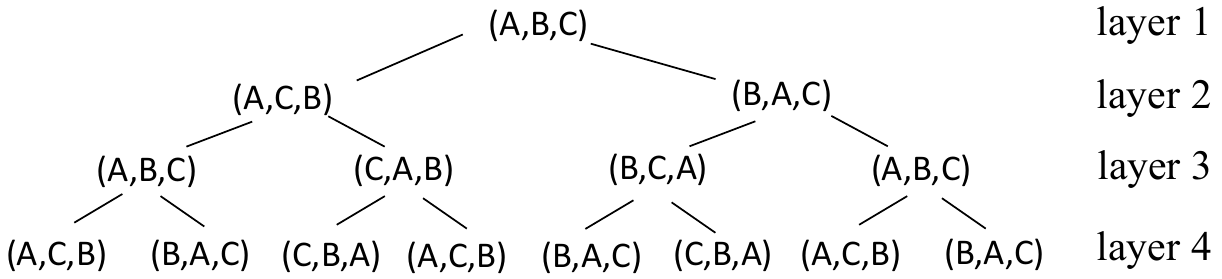}
    \caption{The structure of the move sequence: a binary tree (4 disks)}
  \label{fig:bt}
\end{figure}

The final conclusion is: all individual move can be computed in parallel. In case we want to develop a game from Hanoi Tower, we can compute the next move right away. The formula to compute the $n$th move, when the total number of disks is $N$ and $n=2^k$$\times$$(2m$+$1)$, is given by:\\
\indent $(a,b,c)=(A,B,C)\wedge odd(N$-$k)\oplus(A,C,B)\wedge even(N$-$k)$,\\
and the move is $rotate^m(a, b, c)$, i.e. to rotate $(a, b, c)$ $m$ times. The right side of above equation suggests a new way to write conditional expressions.

For example, for $N=4$ and $n=10$, we have $k=1$ and $m=2$. Since $N-k=3$ is odd, we have $(a, b, c)=(A, B, C)$. To rotate $(A, B, C)$ twice $(m=2)$, obtained is $(B, C, A)$. Thus, the 10th move is to take the top disk on stack $B$ and to put it on top of stack $A$. Readers may test it by hand simulation since there are altogether 15 moves only for $N=4$.

\end{example}

\subsection{Pointers in C}

\subsubsection{Semantic axioms for pointers}\leavevmode

Pointers~\cite{PoC97} are special variables via which other variables can be indirectly accessed. It is useful for call-by-value functions. Extra operations, in addition to read/write, need to be defined for pointers as their due abstraction.

Careful study of the physical facet of pointers leads to a better understanding.

A memory location may be accessed by its address, its variable name and a pointer pointing to it. Address is its hardware name ($hn$ for short) since address is inseparable from hardware. A variable name is its compiler name ($cn$ for short) since it is the compiler that assigns this address to the variable. A pointer is a false name ($fn$ for short) of it since only indirect access is possible. Let be $HN$=$\{hn\}$, $CN$=$\{cn\}$ and $FN$=$\{fn\}$, i.e. $hn$, $cn$ and $fn$ are individuals while $HN$, $CN$ and $FN$ are the respect sets of individuals. Furthermore, $FN^2$ is the pointer set of pointer to pointer.

It is important to notice that $HN$(the set of addresses) is in fact a new kind of data type, since it allows plus and minus operations to be applied as long as the returned value falls into $HN$. In other word, an address is also a two-facet object like a variable.

There are two operations, namely $*$ and $\&$, defined in C on pointers: \\
\indent $*p$=$a$ and $p$=$\&a$ \\
\indent for pointer $p$ pointing to variable $a$. Now, $*p$ is another way to name variable $a$. Let’s call $*p$ anonymous name ($an$ for short) of $a$, since the name $a$ is not explicitly mentioned. Besides, whenever $p$ is assigned a new address, $*p$=$a$ will no longer be true. Let be $AN$=$\{an\}$.

Here, $*$ is a math operator, since its operand is the content of pointer $p$, i.e. a value of type $HN$, a math object; $\&$ is a operator whose operand is the variable name, not the value of this variable, that is a physical object. Thus, as the respect formal definitions of $*$ and $\&$, we have
\begin{align*} 
*: HN \rightarrow CN,\\
\&: CN \rightarrow HN.  
\end{align*}

Note that the definition of $*$ is $HN \rightarrow CN$ instead of $FN \rightarrow CN$, since the content of a pointer (value of type $HN$) is the real operand of $*$. Based on these two definitions, we have the semantics axioms below:\\
Axioms for pointers (AP1-6)\\
(AP1)$\&* = id: HN \rightarrow HN, *\& = id: CN \rightarrow CN$,
\\where $id$ is the identification operator defined respectively on HN and CN.\\
(AP2) $\forall v\in CN : \&v \in HN \wedge constant(\&v) $, once an address is assigned to $v$, it will never change.\\
(AP3) $\&u = \&v \rightarrow u \equiv v$, if two variables share the same address, then they are always equivalent with each other.\\
(AP4) $*p = *q \equiv p = q$, if two different pointers point to the same variable, then they are only equal in value.\\
(AP5)$\overline{p}(\&a)\rightarrow \uline{p} = \&a \wedge *\uline{p} = a$. $*p$ is changed to $*\uline{p}$ by $\overline{p}(\&a)$. \\

Remarks
\begin{itemize}
\item The difference between (AP3) and (AP4) is that $u = v$ will always be true while $*p = *q$ may be changed when $p$ or $q$ receives a new value in $HN$. 
\item (AP5) is an extension of of (A1.1). Here, $*\uline{p} = a$ instead of $\uline{*p} = a$ since $*$ is a math operator. $*p$ is the variable $p$ points to before $\overline{p}(\&a)$. Note that $\&\uline{p}$ is illegal since the operand of $\&$ is a variable name, not a value. 
\end{itemize}

\subsubsection{What is new in the understanding of pointers}\leavevmode

\begin{enumerate}
  \item Here the set of addresses is a new data type for it allows restricted plus and minus operations.
  \item Here it is made clear the difference between physical facet and math facet of a variable while they are called L-value and R-value respectively in the literature on C.
  \item Here the essence of the two unary operators, namely $*$ and $\&$, are made clear: the former is a math operator on $HN$ while the latter is a physical operator on $CN$. For a C compiler, operator $*$ needs the address of $p$ to indirect to where $p$ points to. But for semantic computation, it is not necessary to mention the address of $p$. $\&\uline{p}$ is illegal since the operand of $\&$ is a physical object and $\uline{p}$ is a value.
  \item In case $p = 1000$, $1000 \in HN$, $*1000$ is considered as an illegal L-value by C though the operand of $*$ is an address. OE respects this fact since a C compiler needs a pointer for indirection. Besides, it is unknown whether address 1000 is available to the program.
  \item With axioms (AP1-5), semantics of pointers can be computed from program text as illustrated by examples below. 
\end{enumerate}

\subsubsection{Examples of pointers' semantics computation}\leavevmode

The Grace letter $\Psi$ is used as a general default value for all data types in case a variable is not initialized. $\Psi$ is just used to find absence of initial values in the course of semantic computation, it is not a consisting part of OE.

\begin{example}\label{exp:6-1}
This is from the book \textless Pointer on C\textgreater by Kenneth Reek, 2013, P140\\
\indent Text in C:
\begin{align*}
&int\ a   = 12;\\
&int\ *b  = \&a;\\
&int\ **c = \&b 
\end{align*}
\indent Text in OE:
\begin{align*}
&int\ a, fn\ of\ int\ b,\ fn^2\ of\ int\ c;\\
&\overline{a}(12);\overline{b}(\&a);\overline{c}(\&b);
\end{align*}

To find the values $\uline{a}$, $\uline{b}$ and $\uline{c}$, $*\uline{b}$ and $*\uline{c}$, $**\uline{c}$.

From the text in OE and the axioms above, we have the semantic formula below:\\
$\uline{a} = 12 \wedge \uline{b} = \&a \wedge \uline{c} = \&b$ since 12 and $\&a$ and $\&b$ are constants. So, $*\uline{b} = a \wedge *\uline{c} = b$ by (AP1).

From $\uline{b} = \&a; \uline{c} = \&b$ we have $\uline{b} = \&a; **\uline{c} = **\&b \equiv \uline{b} = \&a; **\uline{c} = *b$. Applying the relay rule in semantic calculus and (AP1), obtained is $**\uline{c} = *\uwave{\ b} \wedge \uwave{\ b} = \&a$ i.e. $**\uline{c} = a$.

Putting above results together:\\
\indent $\uline{a} = 12 \wedge \uline{b} = \&a \wedge \uline{c} = \&b \wedge *\uline{b} = a \wedge *\uline{c} = b \wedge **\uline{c} = a$. \\
\indent As for values of $*\uline{b}$, $*\uline{c}$ and $**\uline{c}$, they are respectively 12, $\&a$ and 12 since different names of the same address share the same value.
\end{example}

\begin{example}\label{exp:6-2}
This example is also from the same book, p137.\\
\indent Text in C:
\begin{align*}
&int\ a;\\
&int\ *d  = \&a;\\
&*d = 10 - *d;\\
&d = 10 - *d 
\end{align*}
\indent Text in OE:
\begin{align*}
&int\ a, fn\ of\ int\ d;\\
&\overline{a}(\Psi);\overline{d}(\&a);\overline{*d}(10-*d);\overline{d}(10-*d)
\end{align*}
\indent To compute $\uline{a}$, $\uline{d}$ and $*\uline{d}$. 

First the semantics of $\overline{a}(\Psi);\overline{d}(\&a);\overline{*d}(10-*d)$:\\
$a = \Psi \wedge \uline{d} = \&a$ is the semantics of $\overline{a}(\Psi);\overline{d}(\&a)$ that provides initial values for $\overline{*d}(10-*d)$ whose semantics is $\uline{*d}=10-*d$. Since the initial value of $d$ is $\&a$, so the semantics of $\overline{a}(\Psi);\overline{d}(\&a)$ is $\uline{*\&a}=10-a$, in which $a =\Psi$ is the initial value, i.e. $\uline{a} = 10 –\Psi$.  Finally, the semantics of $\overline{a}(\Psi);\overline{d}(\&a);\overline{*d}(10-*d)$ is \\
\indent $\uline{d}=\&a\wedge\uline{a}=10-\Psi$ that implies $*\uline{d}=a$.\\
\indent With these as the initial values of the last term $\overline{d}(10-*d)$, we have\\
\indent $\uline{d} = 10 – a = 10 – (10 – \Psi) = \Psi$. Finally $\uline{a} = 10 – \Psi \wedge \uline{d} = \Psi$.

Apparently, there are two errors here: lacking initial value and wrong address.

Note that the process of passing initial values to the next term is just the application of the rule of relay in semantic calculus that can be done in an automatic way by symbolic computation. We didn’t put down all formulas just to make it easier to read.
\end{example}

\section{Conclusion and future work}\label{sec:con}

\begin{figure}
  \centering
 \includegraphics[width=3.2in]{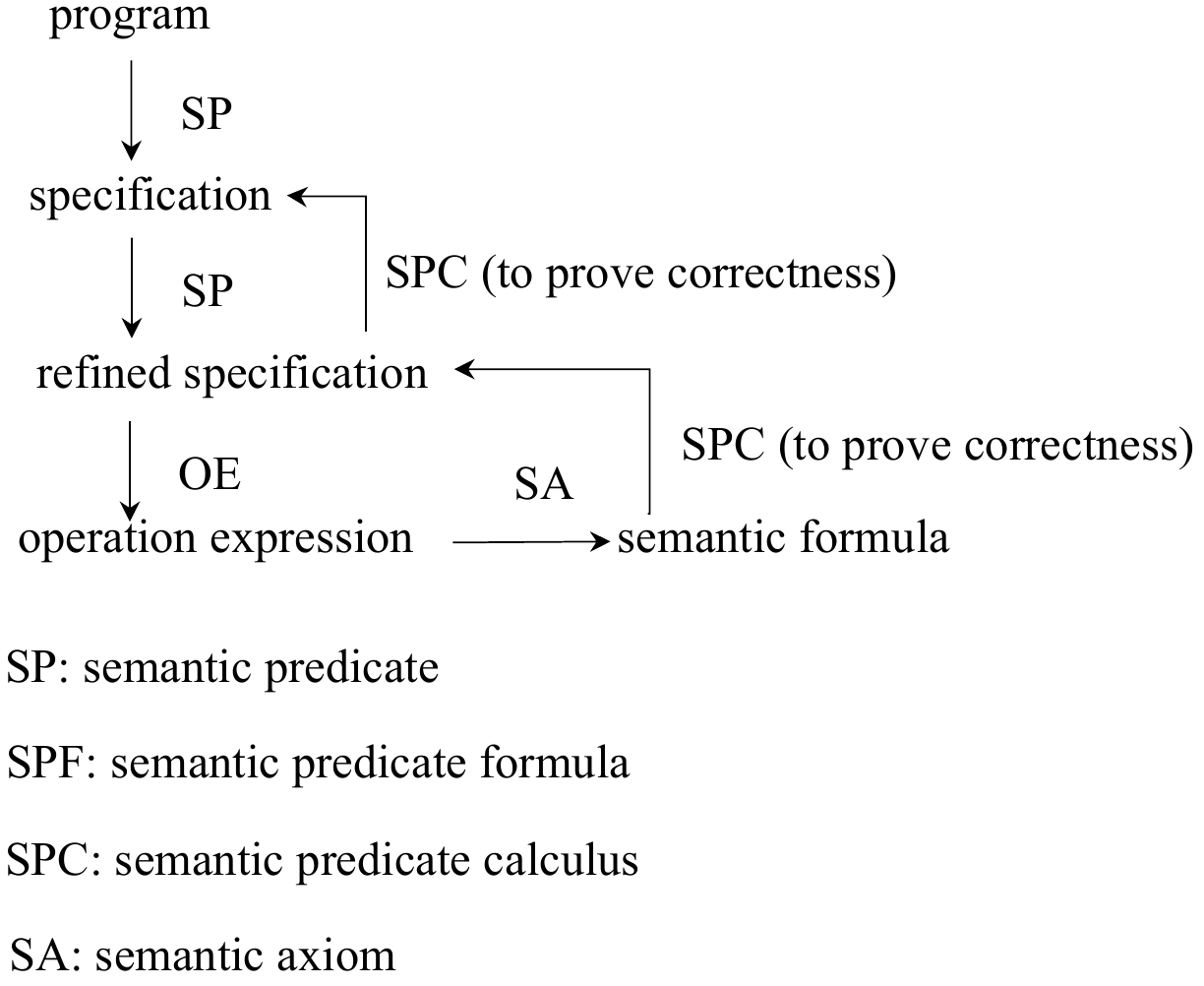}
    \caption{New software lifecycle of OESPA}
  \label{fig:sl}
\end{figure}

Computer science is to lay a solid foundation for programming. OESPA is just one step forward towards this goal. Although OESPA is essentially different from Hoare logic (see Table~\ref{table:com}), it is still not possible to expect programmers to give up what they are familiar in their daily work to switch to OESPA. OESPA is important since it prepares a theoretical foundation for the future generation computer systems. Contributions of OESPA include:

\begin{itemize}
\item[1.] It has formally defined, for the first time in the literature, the concept of program variables (in contrast to the concept of math variables) to serve the need of formal semantics study.
\item[2.] It includes math concepts like semantic functions, semantic predicates, semantic calculus etc. that are badly needed by formal semantics study, but do not exist in conventional mathematics. As shown in Figure~\ref{fig:sl}, these concepts provide means for the analysis of all steps, but testing, of software lifecycle.
\item[3.] It has given program semantics a precise formal definition.
\item[4.] It makes semantic computation possible.
\item[5.] It has clarified the concept of ``computer science''. This term is often mixed up with computational mathematics while the latter belongs to computer application.
\end{itemize}

Real programs are great in size. But this is not a problem at all since a program may be divided into consisting subprogram. Programmers used to insert comments into program text to notify progress. With OESPA, such comments should be replaced by semantic predicate (SP) to make progress explicit. To put all these SP together, obtained is a SP formula, i.e. semantics of the program. Isn’t it nice?

We are well aware that there is still a long way to go for OESPA to play a role in practical programming. To start with in our next research, we would propose rules to interpret, from individual features of C, to semantic predicates. For example, $x$=$e$ is an assignment in C, it is interpreted as $\uline{x}$=$e$. Other programming languages, like Java or Python, may be chosen as well for semantics interpretation. If this is well done for C, a C program can then be interpreted as a semantic formula that complies to semantic calculus. In case this is arranged as part of work agenda, the development of auto tools to carry out the interpretation and to do semantic calculus should also be arranged. Without such auto tools, it is impossible to interpret programs consisting of hundreds of thousands of lines. We hope that programmers may benefit from reading this paper since the above contributions would help in understanding program and programming. You may try it by writing small programs in the way as Figure~\ref{fig:sl} illustrated, An OE program is exactly an expression of operations, no need for key words. As readers can see in this paper, OE programs appear much simpler than conventional ones. The iteration program given above in C has about 10 lines while the corresponding OE program is one line in length. That is why all examples in this paper are very small in size.
 
We do not expect OE to be in practical use before it is enriched (by adapting data types, data structures etc.) to be a programming language and auto tools are available. Our next goal is to have a team to work together to do all these. We know that the bibliography is very short. This is due to the fact that our research is in a completely new direction and as such no similar research reported yet.  Besides, conventional semantics are, in a sense, under criticism here, and it is sufficient to list just a few.

\section*{Acknowledgements}

The authors would like to extend their sincere gratitude to Jianhui Chen for his valuable suggestions. The work was supported by the National Key Research and Development Program of China (No. 2019YFB1704003), the National Nature Science Foundation of China (No. 71690231) and Tsinghua BNRist.

\bibliographystyle{unsrt}
\bibliography{ref}

\begin{thebibliography}{10}

\bibitem{T76}
R.~D. Tennent.
\newblock The denotational semantics of programming languages.
\newblock {\em Communications of the ACM}, 19(8):437--453, 1976.

\bibitem{P81}
G.~D. Plotkin.
\newblock {\em A structural approach to operational semantics}.
\newblock Aarhus university, 1981.

\bibitem{H69}
C.~A.~R. Hoare.
\newblock An axiomatic basis for computer programming.
\newblock {\em Communications of the ACM}, 12(10):576--580, 1969.

\bibitem{HH98}
J.~He C.~A.~R.~Hoare.
\newblock {\em Unifying theories of programming}.
\newblock Englewood Cliffs: Prentice Hall, 1998.

\bibitem{D76}
E.~W. Dijkstra.
\newblock {\em A discipline of programming}.
\newblock Englewood Cliffs: Prentice Hall, 1976.

\bibitem{G81}
D.~Gries.
\newblock {\em The science of programming}.
\newblock Springer Science \& Business Media, 1981.

\bibitem{CM88}
J.~Misra K.~M.~Chandy.
\newblock {\em Parallel Program Design: A Foundation}.
\newblock Addison-Wesley publishing company, 1988.

\bibitem{CW96}
J.~M.~Wing E.~M.~Clarke.
\newblock Formal methods: state of the art and future directions.
\newblock {\em ACM Computing Surveys}, 28(4):626--643, 1996.

\bibitem{QS61}
C.~A.~R. Hoare.
\newblock Algorithm 64: quicksort.
\newblock {\em Communications of the ACM}, 4(7):321--322, 1961.

\bibitem{OESPA}
C.~Yuan.
\newblock {\em OESPA: Semantic oriented theory of programming}.
\newblock Science Press, China, 2019.

\bibitem{PoC97}
K.~A. Reek.
\newblock {\em Pointers on C}.
\newblock Addison-Wesley Longman Publishing Co., Inc., 1997.

\end{thebibliography}

\end{document}